\documentclass[draft]{agujournal2019}
\usepackage{url} 
\usepackage{lineno}
\usepackage[inline]{trackchanges} 
\usepackage{soul}

\usepackage{xcolor}
\newcommand{\notes}[1]{{\color{red} #1}}

\draftfalse

\journalname{Journal of Advances in Modeling Earth Systems}

\begin{document}

%
%

\title{Controlled abstention neural networks for identifying skillful predictions for classification problems}

%
%
\authors{Elizabeth A. Barnes\affil{1} and Randal J. Barnes\affil{2}}
\affiliation{1}{Department of Atmospheric Science, Colorado State University, Fort Collins, CO, USA.}
\affiliation{2}{Civil, Environmental, and Geo- Engineering, University of Minnesota, Minneapolis, MN, USA.}




\correspondingauthor{Elizabeth A. Barnes}{eabarnes@rams.colostate.edu}




\begin{keypoints}
\item A simple neural network approach for abstention is explored for climate classification problems
\item A new abstention loss is introduced to identify, and preferentially learn from, more confident samples
\item This new abstention loss improves prediction accuracy for a variety of climate use cases
\end{keypoints}



\begin{abstract}
The earth system is exceedingly complex and often chaotic in nature, making prediction incredibly challenging: we cannot expect to make perfect predictions all of the time. Instead, we look for specific states of the system that lead to more predictable behavior than others, often termed ``forecasts of opportunity.'' When these opportunities are not present, scientists need prediction systems that are capable of saying ``I don't know.'' We introduce a novel loss function, termed the ``NotWrong loss'', that allows neural networks to identify forecasts of opportunity for classification problems. The NotWrong loss introduces an abstention class that allows the network to identify the more confident samples and abstain (say ``I don't know'') on the less confident samples. The abstention loss is designed to abstain on a user-defined fraction of the samples via a PID controller. Unlike many machine learning methods used to reject samples post-training, the NotWrong loss is applied during training to preferentially learn from the more confident samples.  We show that the NotWrong loss outperforms other existing loss functions for multiple climate use cases. The implementation of the proposed loss function is straightforward in most network architectures designed for classification as it only requires the addition of an abstention class to the output layer and modification of the loss function. 
\end{abstract}

\section*{Plain Language Summary}
The earth system is exceedingly complex and often chaotic in nature, making prediction incredibly challenging: we cannot expect to make perfect predictions all of the time. Instead, we can look for specific states of the system that lead to more predictable behavior than others, often termed ``forecasts of opportunity''. When these opportunities are not present, scientists need prediction systems that are capable of saying ``I don't know.'' We present a method for teaching neural networks, a type of machine learning tool, to say ``I don't know'' for classification problems. By doing so, the neural network focuses less on the predictions it identifies as problematic and focuses more on the predictions where its confidence is high. In the end, this leads to better predictions.

%
%
\clearpage

\section{Introduction}
The earth system is exceedingly complex and often chaotic in nature, making prediction incredibly challenging: we cannot expect to make perfect predictions all of the time. Instead, we look for specific states of the system that lead to more predictable behavior than others, often termed ``forecasts of opportunity''  \cite{Mariotti2020,Albers2019,Mayer2020,Barnes2020}. When skillful forecast opportunities are not present, scientists need prediction systems that are capable of saying ``I don't know.'' While this concept of forecasts of opportunity stems from weather and climate predictions, the idea is more general than this. For example, a forecast of opportunity framework may be beneficial when certain predictors are only helpful under certain circumstances. Additionally, if certain predictions (labels) are more predictable than others, or if there is unstructured noise in the training data, a system that can say ``I don't know'' may identify the more skillful predictions, when they occur.

Many approaches to identify skillful forecasts of opportunity already exist. For example, retrospective analysis of the forecast can provide a sense of the physical circumstances that can lead to forecast successes or busts \cite<e.g.>{Rodwell2013-qz}, The ensemble spread can also give a sense of uncertainty in numerical weather prediction systems \cite<e.g.>{Van_Schaeybroeck2016-lo}. \citeA{Albers2019} used a linear inverse modeling approach to identify confident subseasonal predictions and showed that these more confident predictions were indeed more skillful. Recently, \citeA{Mayer2020} and \citeA{Barnes2020} suggested that machine learning, specifically neural networks, may be a useful tool to  identify forecasts of opportunity for subseasonal-to-seasonal climate predictions. Specifically, a classification network is first trained, then the predicted probabilities are ordered from largest to smallest. A selection of predictions with the highest probabilities are identified as possible forecasts of opportunity. While \citeA{Mayer2020} and \citeA{Barnes2020} show that this approach works well for classification tasks (i.e., predicting a specific category) where the network is already tasked with predicting a probability, it is less clear how one might apply this methodology to regression tasks (i.e., predicting a continuous quantity).

Most of the current machine learning approaches used to identify forecasts of opportunity, including those described above, are applied post-training. The network is first trained, and then the model confidence is assessed. Instead, here we lean heavily on work by \citeA{Thulasidasan2019} and \citeA{Thulasidasan2020} to further explore a deep learning abstention loss function for classification tasks that teaches the network to say ``I don't know'' (abstain) on certain samples \textit{during training}. The resulting controlled abstention network (CAN) preferentially learns from the samples in which it has more confidence and abstains on samples in which it has less confidence. The CAN is designed to abstain on a user-defined fraction via a PID controller, which ultimately leads to more accurate predictions than our baseline classification approach. While alternative methods have recently been suggested for abstention (rejection) during training \cite{Geifman2019-paper,Geifman2019-thesis}, the CAN approach can be easily implemented in most any network architecture designed for classification, as it only requires the addition of an abstention  class to the output layer and modification of the training loss.

We demonstrate the behavior of the CAN for three use cases based on synthetic climate data where the correct answer is known. The first use case explores the utility of the CAN in situations where certain classes (labels) are more predictable than others. The second use case explores the ability of the CAN to learn in the presence of unstructured noise, that is, when there is no way to tell \textit{a priori} whether the sample is predictable or not. The third use case is modeled loosely after global teleconnections associated with the El Ni\~no Southern Oscillation \cite<e.g.>{McPhaden2006-pi,Yeh2018-tf} and explores the utility of the CAN in identifying forecasts of opportunity for climate prediction applications.

Section 2 introduces the synthetic climate data and general neural network architecture. Section 3 discusses the baseline loss function and the CAN in detail, and Section 4 presents the results. Additional discussion on the approach compared to previous approaches is provided in Section 5 and conclusions in Section 6.

\section{Data and use cases}
\subsection{Synthetic climate data}
To demonstrate the utility of the controlled abstention network (CAN), we use the synthetic benchmark data set introduced by \citeA{Mamalakis2021}. While \citeA{Mamalakis2021} provides an extensive description of this data, we give a brief overview here. The dataset consists of input fields $x_i$ and output series $y_i$ (where $i$ denotes the $i^{th}$ sample), which is a function of the input. The input fields represent monthly anomalous global sea surface temperatures (SSTs) generated from a multivariate normal distribution with a correlation matrix estimated from observed SST fields\footnote{https://psl.noaa.gov/data/gridded/data.cobe2.html}. The $i^{th}$ input sample consists of one map of SST anomalies, denoted as $x_i$. \citeA{Mamalakis2021} then define the global response $y_i$ to sample $x_i$ as the sum of local, nonlinear responses. Specifically, 
 \begin{linenomath*}
 \begin{equation}
 y_i = \sum_g F_g(x_i)
 \end{equation}
 \end{linenomath*}
 where $g$ represents the grid point and $F_g$ is defined locally at each grid point $g$ by a piecewise linear function. The slopes $\beta_n$ (where $n$ is an integer that runs from 1 to the number of piecewise linear segments, set here to 5) of each local function are chosen randomly from a multivariate normal distribution with correlation matrix, once again, estimated from observed SST fields. 
 
 In the end, this data set consists of input maps of SSTs with spatial correlations indicative of observed SSTs, but where each input map is independent of the others. $y_i$ then represents the sum of contributions from each grid point across the globe, where that contribution is a nonlinear function (specifically piecewise linear) of the SST value at that grid point. To speed up training time, we reduce the number of grid points (pixels) from that used by \citeA{Mamalakis2021} to 60 longitudes and 15 latitudes for a total of 900 grid points per input map. An example input map is shown in Fig. \ref{fig_arch}.

\subsection{Experimental design}
We modify the synthetic climate data to make it suitable for classification by assigning each input map (i.e. sample) to one of $k=10$ classes (see Fig. \ref{fig_arch}). The ten classes are determined by binning the $y$ values into deciles. For all use cases explored, we task a neural network with ingesting a sample input map and predicting the correct class.

\subsubsection{badClasses}
For the first use case, badClasses, we modify the data set such that all samples in classes 4 and 5 are ``corrupted"; they are randomly assigned an incorrect class. Over the entire data set, 20\% of the samples are assigned incorrect labels and 80\% of the samples retain their correct labels. In this situation, we would like to to see the CAN identify the samples associated with classes 4 and 5 and abstain on them, since any perceived relationship is unreliable. 

\subsubsection{mixedLabels}
For the mixedLabels use case, we modify the data set such that 5\% of all samples (no matter the class) are corrupted by assigning a randomly chosen incorrect label while 95\% of the samples retain their correct label. Unlike badClasses, in this case there is no systematic relationship between the input maps and whether the sample is corrupted or not.  For mixedLabels, we would like the CAN to learn to abstain on the corrupted samples by identifying them as those that do not behave like the majority of the samples. If the CAN does this, it should be able to learn the reliable samples better. For the testing set, the CAN will not know which samples have been corrupted and which have not, but it should achieve better accuracy as it reduced the chances on learning spurious relationships during training.

\subsubsection{fooENSO}
\begin{figure}
\begin{center}
\noindent\includegraphics[width=400px]{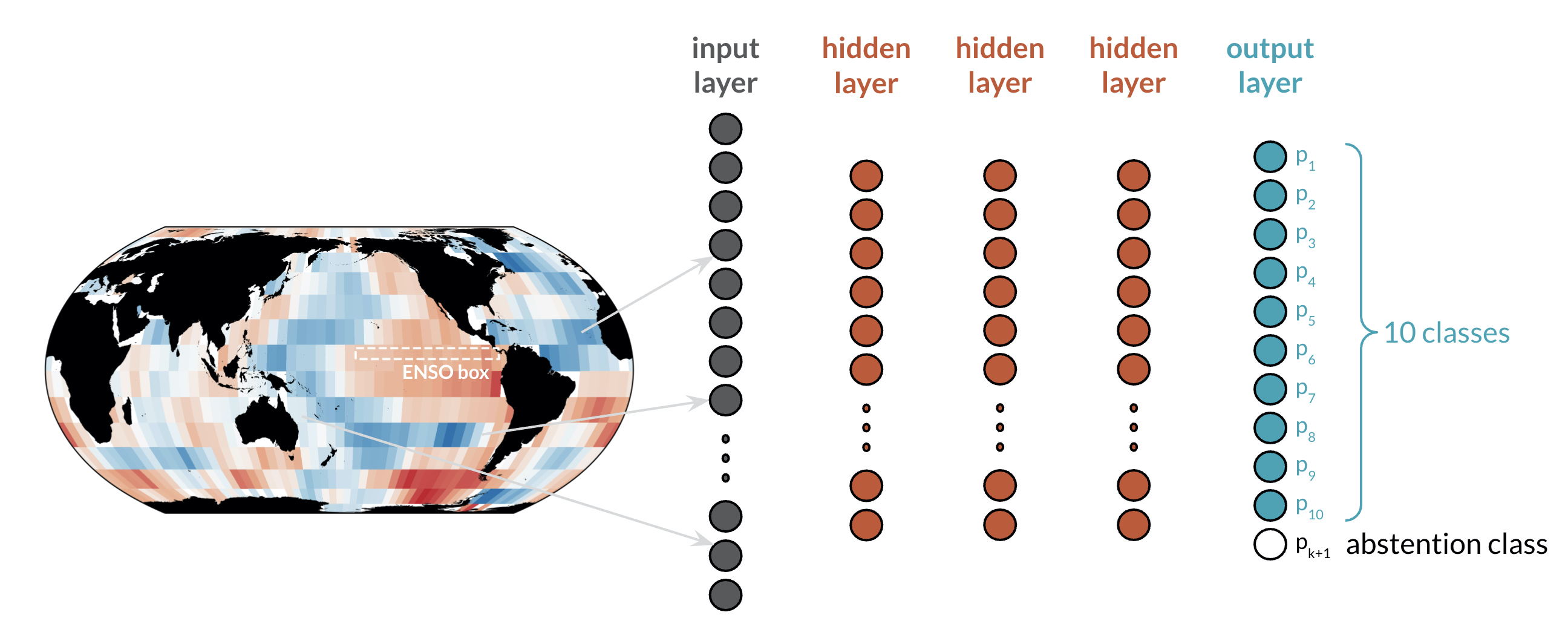}
\end{center}
\caption{General CAN network architecture used for all use cases. A map of synthetic sea-surface temperatures is fed into a fully connected network and tasked with classifying each sample into one of ten classes. An additional class, the abstention class, is included in the output layer for training with abstention. The number of hidden layers varies between two and three depending on the use case.}
\label{fig_arch}
\end{figure}

For our last use case, fooENSO, we modify the data to loosely reflect forecasts of opportunity related to teleconnections associated with the El Ni\~no Southern Oscillation (ENSO). Warm ENSO events (El Ni\~no events) have long been known to impact global temperatures and precipitation \cite<e.g.>{McPhaden2006-pi,Yeh2018-tf}, at times leading to skillful forecasts on subseasonal-to-seasonal time scales \cite<e.g.>{Johnson2014-fh}. To mimic this behavior with our synthetic data set, we average the anomalous synthetic SSTs in the ENSO region within the equatorial eastern Pacific (dashed white box in the map in Fig. \ref{fig_arch}). When the average value in this box is larger than 0.5 (29\% of the samples), we leave the sample as is to reflect an opportunity where a strong El Ni\~no may lead to more predictable behavior of the global climate system. Samples where the average value is less than 0.5 represent ``noisy'' samples; therefore, we corrupt 50\% of these noisy samples by assigning them a randomly chosen incorrect label. As a result, 35\% of all samples are corrupt, and 29\% of all samples (i.e. those associated with a strong El Ni\~no) retain their correct label. We note that one could corrupt all of the noisy samples, instead of just 50\% of them, and the conclusions are the same (see Supp. Fig. S1). With such a setup, we would like the CAN to identify strong synthetic El Ni\~no samples (i.e. large values within the ENSO box, Fig. \ref{fig_arch}) as reliable samples to learn, while abstaining on the other samples that may have unreliable labels.

\subsection{Network architecture and training}
We train a fully connected feed-forward network with two to three hidden layers using Python 3.7.9 and TensorFlow 2.4. For badClasses and mixedLabels, the network has two hidden layers with 50 and 25 units, respectively. For fooENSO, we train a network with three hidden layers with 500, 250, and 20 units, respectively, to demonstrate the utility of the CAN for deep networks. Additional use cases are provided in Supp. Fig. S1. For the baseline ANN, the output layer consists of 10 units, representing each of the 10 classes (Fig. \ref{fig_arch}). For the CAN, we add an additional output unit to represent the abstention class. This will be discussed further in the following subsection. 

We train with a ReLU (rectified linear unit) activation function on the hidden layers and a softmax layer at the output. The softmax layer ensures that the sum of all predicted likelihoods is 1.0 for each sample. The network is trained with a learning rate of 0.001 and batch size of 32. For badClasses and mixedLabels, we train on 8,000 samples, validate on 5,000 samples, and test on 5,000 samples. While we could train on a much larger data set, we have intentionally kept the sample size relatively small to demonstrate the utility of the CAN when the sample size is relatively low --- as is the case for many geoscience applications. For fooENSO, we train on 32,000 samples, validate on 5,000 samples, and test on 5,000 samples. The number of samples is increased for this use case to accommodate the increase in complexity of the network. That said, additional use cases with smaller training sizes are shown in Supp. Fig. S1. All quantities and figures are computed from the testing data unless otherwise specified.

We employ early stopping to automatically determine the optimal number of epochs to train. Specifically, the network stops training when the validation accuracy stops increasing, with a {\tt patience} of 30 epochs. The network with the best performance on the validation accuracy is saved. Specifically for the CAN, we select the best-performing network from epochs where the validation abstention fraction is within 0.1 of the user-chosen abstention setpoint. For all examples shown here, 50 different networks are trained for each configuration (i.e. baseline ANN and CAN) by varying the randomly initialized weights.

\section{Methods}
\subsection{Baseline networks}
The baseline artificial neural network (baseline ANN) is identical to the CAN architecture (Fig. \ref{fig_arch}) with the following exception: it does not include an abstention class, and it uses a standard cross-entropy loss function \cite<>[p. 149]{Geron2019}, which we define as
\begin{equation}
    \mathcal{L}_C(x_{i,j}) = -\log{p_{i,j}}
\end{equation}
where $x$ denotes sample $i$ with true label $j$ (where $1\le j \le k$), and $p_{i,j}$ is the likelihood assigned to the correct class for sample $i$. From this point on we drop the subscript $i$ for readability. 

Once the baseline ANN is trained, we invoke abstention on the least-certain predictions by thresholding the ANN-predicted likelihoods (i.e., the output of the network). Specifically, the class prediction by the network for a single sample is defined as the class with the highest likelihood. We then sort these winning likelihoods to threshold the ANN predictions. For example, we abstain on 20\% of the samples by throwing out the the 20\% smallest winning likelihoods (20\% least confident predictions). This thresholding approach for abstention has been shown to be very powerful on its own \cite<e.g.>{Mayer2020,Barnes2020}, and will serve as a comparison for the CAN.

Throughout this paper, we use ``coverage'' to denote the fraction of samples for which the network makes a prediction, and ``abstention'' to refer to the fraction of samples for which the network does not make a prediction. Thus, the percent coverage is always 100\% minus the percent abstention. For the baseline approach, abstention and coverage is computed post-training based on the predicted winning likelihoods, while for the CAN these quantities are determined during the training itself (see next section).

\subsection{Controlled Abstention Network (CAN)}
\subsubsection{NotWrong loss}
We next introduce the abstention loss function for the controlled abstention network (CAN). Unlike the baseline ANN, the CAN architecture allows the network to assign a label of \textit{abstain}. The abstention loss is designed to penalize the network for abstention, but penalizes the network even more for getting it wrong. For this reason, we have named it the NotWrong loss and define it as
\begin{equation}
    \mathcal{L}_{NW}(x_j) = -\log{\left(p_j+p_{k+1}\right)} - \alpha \log{q} \label{notwrongloss}
\end{equation}
where $p_{k+1}$ is the likelihood assigned to the abstention class, $\alpha$ is a non-negative weight, and $q$ is the likelihood assigned to not abstaining,
\begin{equation}
    q = 1 - p_{k+1}=\sum_{m=1}^kp_m. \label{q_def}
\end{equation}
The first term in Eq. \ref{notwrongloss} represents the likelihood of not getting the prediction wrong, that is, the sum of the likelihood of getting it correct plus the likelihood of abstaining, while the second term is a penalty term for abstaining that is weighted by $\alpha$. Without this penalty the network would abstain on every sample to minimize the loss.

Like the abstention loss introduced in \citeA{Thulasidasan2019}, the NotWrong loss has the property that during gradient descent, the network continues to learn on the abstained samples, although to a lesser extent. A proof of this is provided in the supplemental material. This feature of the loss allows the network to move samples in and out of abstention during training while it continues to learn the non-abstained samples better (as will be shown).

\subsubsection{PID controller}
\begin{figure}
\begin{center}
\noindent\includegraphics[width=195px]{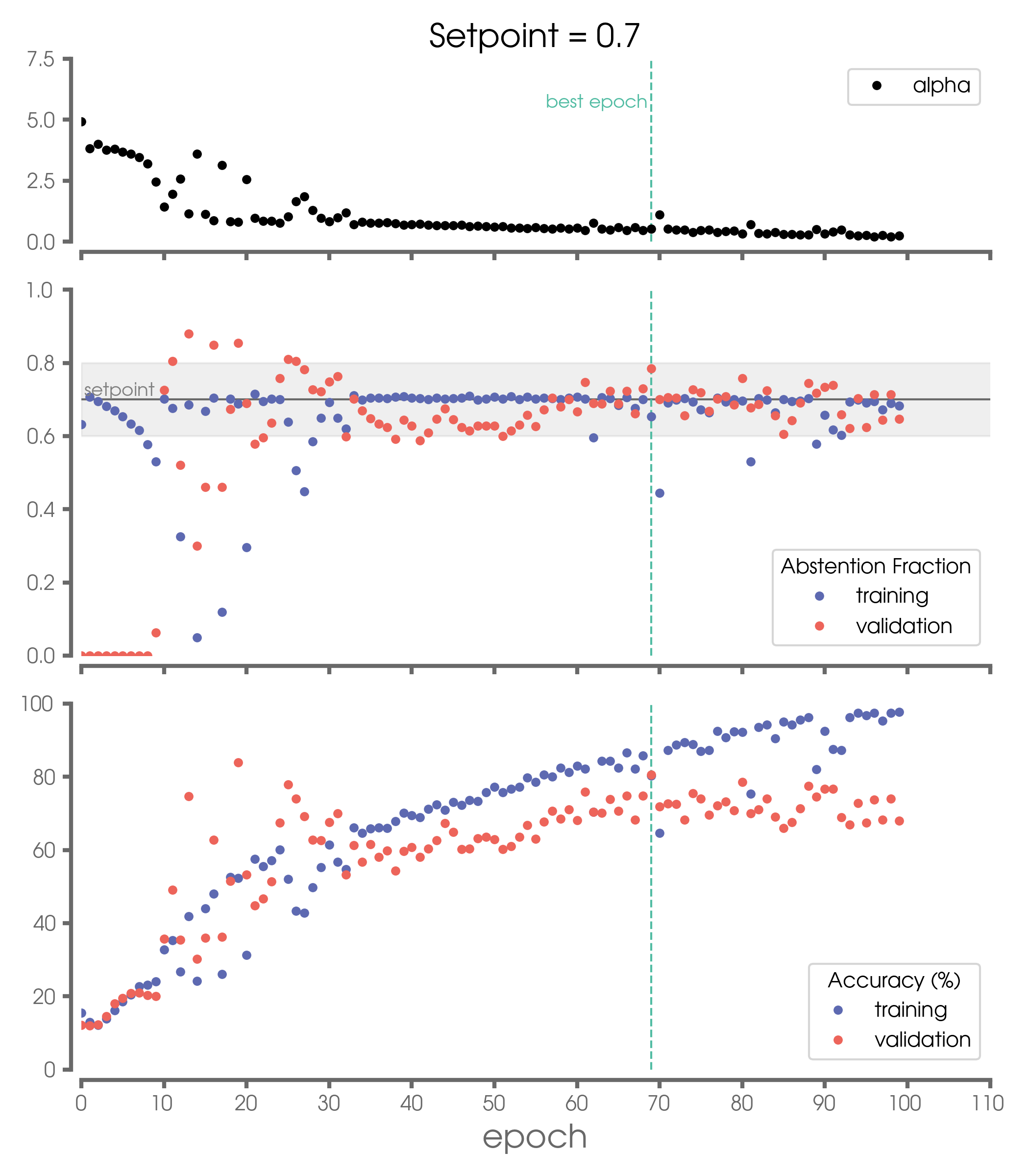}
\end{center}
\caption{Example of CAN metrics during training of the badClasses experiment with abstention setpoint of 0.7 and a network random seed of 19.}
\label{fig_epochs}
\end{figure}

The parameter $\alpha$ in Eq. \ref{notwrongloss} determines how much the network is penalized for abstaining. $\alpha$ can be adaptively modified throughout training so that the network abstains on a specified percent of the training samples. Inspired by the success reported in \citeA<>[Chapter 4]{Thulasidasan2020}, we implement a discrete-time PID controller (velocity algorithm) to modulate $\alpha$ throughout training \cite<e.g,>[Eq. (1.38)]{Visioli2006}. 

\citeA{Thulasidasan2020} solely explores low-abstention setpoints (e.g. 10\%), and evaluates the PID terms batch by batch. For our applications, however, we need the algorithm to work well for a broad range of abstention setpoints (e.g. from 10\% to 90\%). With a high abstension setpoint, say 90\%, and a batch size of 32, only 3 samples on average would be covered per batch --- this leads to to unstable behavior. Because of this, we evaluate the PID terms on 6 consecutive batches ($32 \times 6 = 192$ samples); this strategy leads to more stable behavior of the abstention fraction, but it does not impede training. An example where the PID controller modulates $\alpha$ to control the abstention setpoint during training is shown in Fig. \ref{fig_epochs}a,b.

\section{Results}
\subsection{General performance}
\begin{figure}
\begin{center}
\noindent\includegraphics[width=275px]{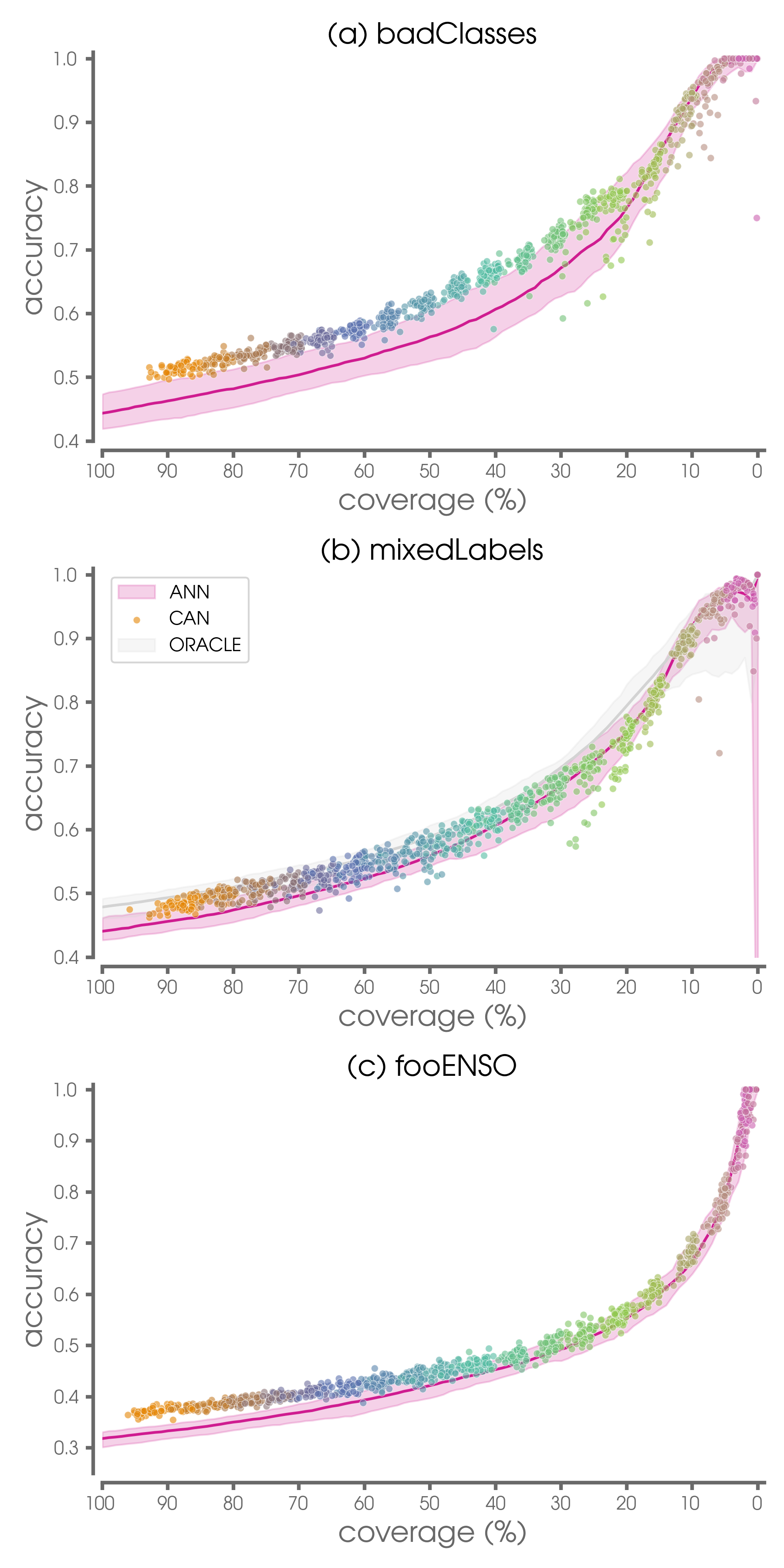}
\end{center}
\caption{CAN testing accuracy as a function of percent coverage for 50 different simulations of each of the three use cases. Pink shading denotes the full range of accuracies of the baseline ANN, while the solid pink line denotes their median. Dots denote results for each CAN simulation; colors denote different abstention setpoints. Panel (b) includes an additional gray line that shows the full range of results from the ORACLE simulation.}
\label{fig_acc}
\end{figure}

\begin{figure}
\begin{center}
\noindent\includegraphics[width=350px]{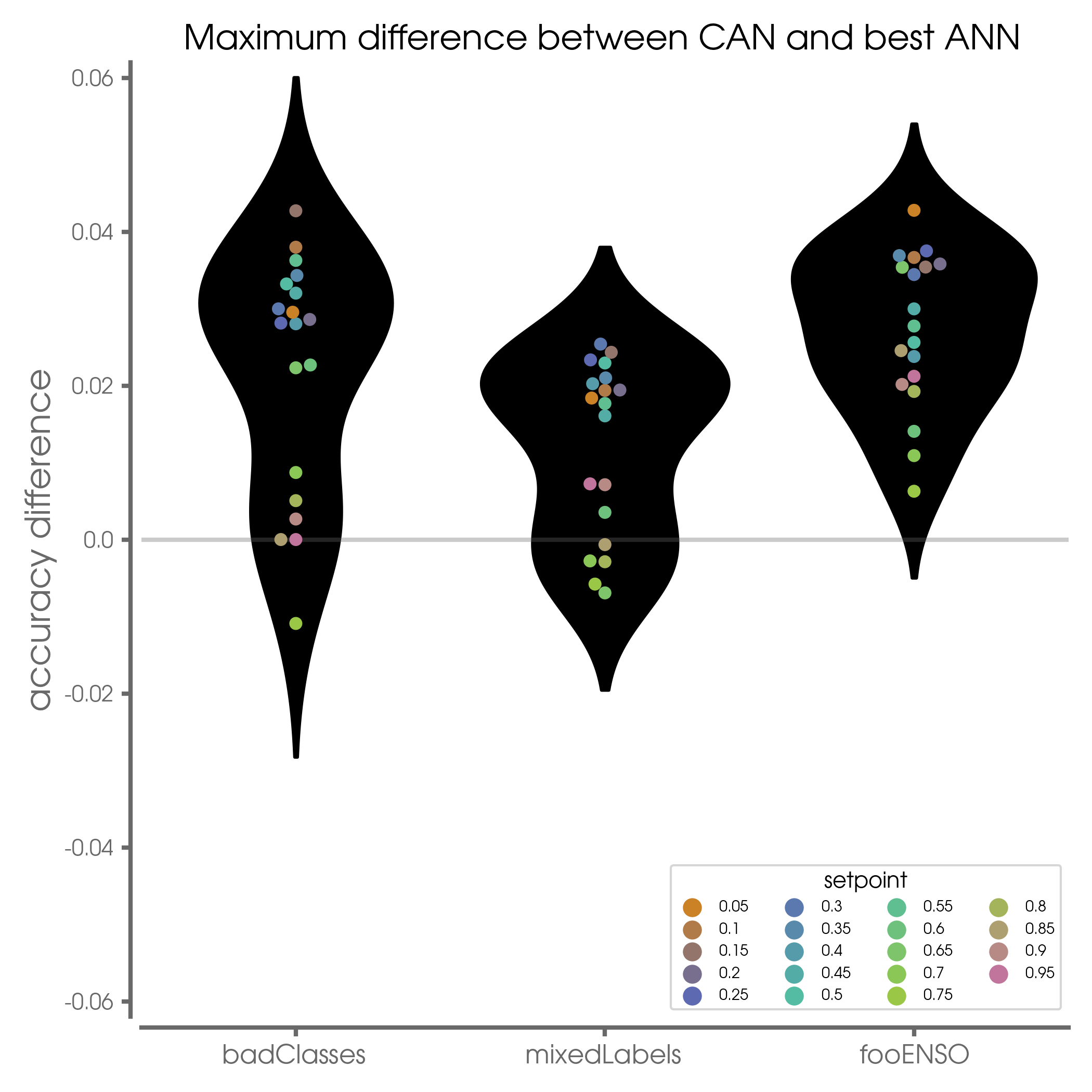}
\end{center}
\caption{Maximum difference in testing accuracy between the CAN and the best baseline ANN for various abstention setpoints. Positive values imply higher accuracies of the CAN compared to the best baseline ANN.}
\label{fig_max}
\end{figure}

For each use case, we train 50 different baseline ANNs and 50 different CAN networks for setpoints ranging from 0.05 to 0.95. Accuracies as a function of percent coverage for the three use cases are shown in Fig. \ref{fig_acc}. As coverage decreases (abstention increases) accuracy increases for the baseline ANNs and CANs. This demonstrates that more confident predictions are more accurate. That said, for all three use cases, the CAN exhibits higher accuracies compared to the baseline ANN for most coverages. This is further visualized in Fig. \ref{fig_max}, where we plot the maximum difference in accuracy between the CAN and the best baseline ANN for the same coverage. For most abstention setpoints the CAN shows an improvement on the baseline ANN. These improvements can be as high as an increase in accuracy of 0.045 (i.e. 4.5\%).

For the $mixedLabels$ use case, Fig. \ref{fig_acc}b shows an additional gray shaded line that represents an additional set of 50 simulations we term ORACLE \cite<see also>{Thulasidasan2020}. In this setup, we play an all-knowing oracle and remove all of the corrupted samples from the training and validation sets prior to training using the baseline ANN approach. Accuracies are then evaluated on the same testing set as the CAN. Thus, ORACLE represents an upper bound on what we could hope to expect from the CAN. For coverage fractions between 40\%-80\%, the best CAN models achieve similar accuracies to ORACLE. This suggests that the abstention process has done a nearly ideal job abstaining on the corrupted samples.

\subsection{Strategies of the CAN}

\begin{figure}
\begin{center}
\noindent\includegraphics[width=400px]{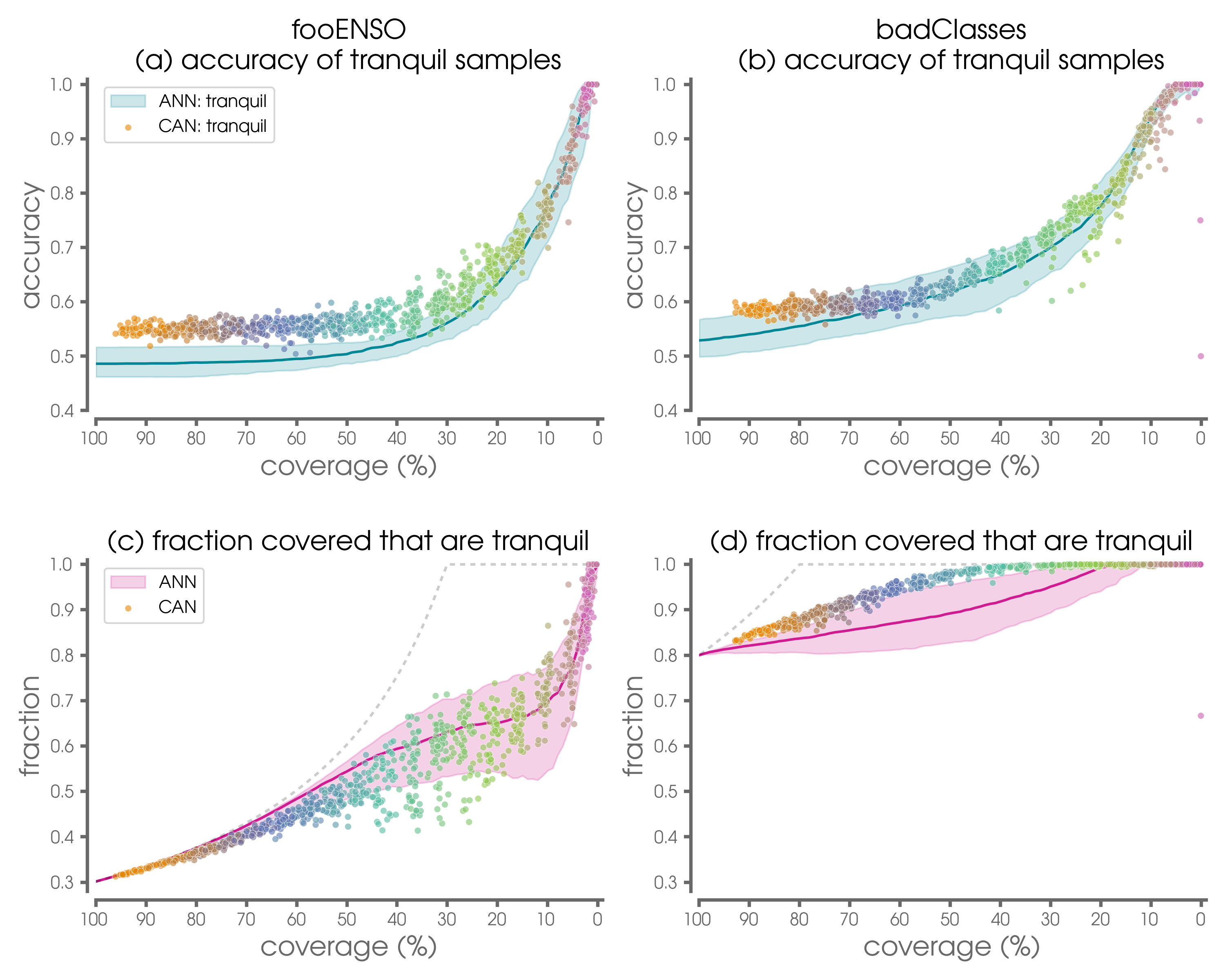}
\end{center}
\caption{(a,b) Testing accuracy computed over the tranquil samples for the fooENSO and badClasses experiments. Shading denotes the full range of accuracies from the baseline ANN, while the solid lines denote the median. (c,d) Fraction of covered samples that are tranquil samples. Gray dashed lines denote the maximum fraction possible given the experimental setup.}
\label{fig_approach}
\end{figure}

There are multiple strategies that the CAN may employ to improve upon the baseline ANN: (1) the CAN may do a better job identifying the non-corrupted samples and abstaining on the corrupted samples, (2) the CAN may do a better job learning the relationship between the non-corrupted inputs and their labels/true classes, and (3) a combination of \#1 and \#2. Fig. \ref{fig_approach} explores these strategies for the badClasses and fooENSO use cases. 

Beginning with the fooENSO use case (Fig. \ref{fig_approach}a,c), we once again plot the accuracy as a function of coverage, but focus only on the tranquil samples this time. We use the identifier \textit{tranquil} to denote the group of samples that should be identifiable by the network as non-corrupt (i.e. all samples that exhibit a strong El Ni\~no) and \textit{not tranquil} to denote the rest. Fig. \ref{fig_approach}c shows that while the baseline ANN and the CAN cover similar fractions of tranquil samples, the CAN is better at learning the relationship between the inputs and outputs of the tranquil samples (Fig. \ref{fig_approach}a). Thus, for this use case, the CAN outperforms the baseline ANN by taking approach \#2. For the badClasses use case (Fig. \ref{fig_approach}b,d), the CAN appears to outperform the baseline ANN by employing a combination of approaches \#1 and \#2. That is, the CAN does a significantly better job identifying the tranquil samples (i.e. samples that do not belong to classes 4 and 5; Fig. \ref{fig_approach}d). It also does a better job learning the relationship between the inputs and outputs of the tranquil samples.

\begin{figure}
\begin{center}
\noindent\includegraphics[width=300px]{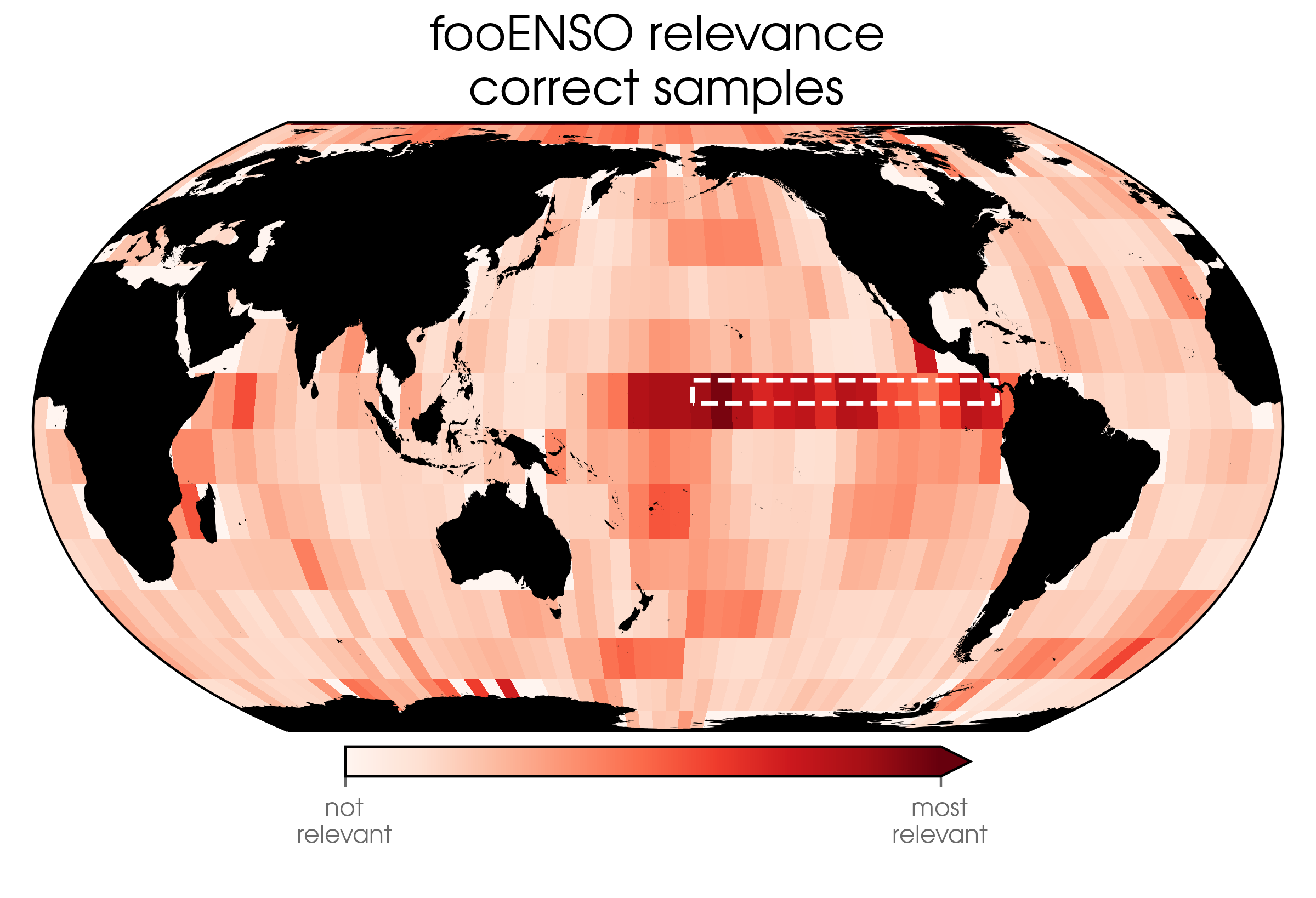}
\end{center}
\caption{Layer-wise propagation (LRP) relevance heatmap averaged over testing samples for the fooENSO experiment using an abstention setpoint of 0.7. The heatmap shows the mean over all 50 simulations for samples that are correctly assigned a label of 1 (i.e. $j=1$). We use the LRP-z rule but set all negative relevances to zero and scale each heatmap by the sum of the positive relevances over the map prior to averaging.}
\label{fig_lrp}
\end{figure}

Neural network explainability methods have recently attracted the attention of the climate science community to assist scientists in identifying new forecasts of opportunity  \cite<e.g>{Toms2020,Barnes2020,Mayer2020}. One particular method, layer-wise relevance propagation (LRP), produces a heatmap which approximates the most relevant regions for a neural network's output according to a set of propagation rules \cite<e.g.>{Bach2015,Montavon2017,Mamalakis2021}. 
Fig. \ref{fig_lrp} shows the average fooENSO LRP heatmap over all 50 simulations for samples that were correctly assigned a label of 1 (i.e. $j=1$) by the CAN. Large relevance within the ENSO region (dashed white box) demonstrates that the CAN is indeed using this region to identify successful predictions. Other regions are also relevant (non-zero), since the correct label $j=1$ is determined by the sum of contributions across the globe (see Section 2.2.3). In this way, the CAN architecture may be paired with explainability methods to identify the mechanisms behind forecasts of opportunity within data sets \cite<e.g.>{Barnes2020}. Likewise, one could also use explainability methods to explore the reasons behind the CAN's abstention \cite<e.g.>{Thulasidasan2019}.

\section{Discussion}
As discussed in the introduction, \citeA{Thulasidasan2019} and \citeA{Thulasidasan2020} introduced their own deep abstention classifier (DAC) loss function, defined as
\begin{equation}
    \mathcal{L}_{DAC}(x_j) = -q\log{\left( \frac{p_j}{q }\right)} - \alpha \log{q} \label{dacloss}
\end{equation}
where all notation is the same as in Eqs. \ref{notwrongloss} and \ref{q_def}. Our NotWrong loss and the DAC loss are very similar. The important difference is that the quantity inside the first log in the DAC loss represents the likelihood of getting the prediction \textit{correct}, while in the NotWrong loss it represents the likelihood of getting the prediction \textit{not wrong}. 

While both the DAC loss and NotWrong loss improve accuracies over the baseline ANN, for the use cases explored here we find that the NotWrong loss outperforms the DAC loss, as shown in Supp Fig. S2. We believe that this is because the NotWrong loss puts more energy into learning the correct answer via larger negative derivatives of the loss with respect to the correct class ($a_j$). This is demonstrated in Supp. Fig. S3, where we display the derivative of the loss with respect to $a_j$ for different value ranges of $p_j$ and $p_{k+1}$. Pink shading in Supp. Fig. S3c represents regions in phase space where the derivative of the NotWrong loss is more negative than the DAC loss, and we find this region of phase space most representative of the use cases explored here (not shown).  Future work will explore this behavior further. 

\section{Conclusions}
The ability to say ``I don't know'' is an important skill for any scientist. 

In the context of prediction with deep learning, the identification of uncertain (unpredictable) samples is often approached post-training. In this paper, we explore an alternative: a deep learning loss function that can abstain \textit{during training} for classification problems. We introduce a new abstention loss for classification that focuses on getting the answer not wrong, rather than getting it right. The controlled abstention network (CAN) with this loss allows the network to preferentially learn more from confident samples, and ultimately outperform both the baseline ANN approach and the abstention loss of \citeA{Thulasidasan2019} and \citeA{Thulasidasan2020} for the climate use cases explored here. 

An additional benefit of the abstention loss CAN is its simplicity --- it is straightforward to implement in most any network architecture, as it only requires adding an additional class to the output layer and modification of the training loss. The abstention loss framework has the potential to aid deep learning algorithms to identify skillful forecasts, which ultimately improves performance on the samples with predictability.

\acknowledgments
This work was funded, in part, by the NSF AI Institute for Research on Trustworthy AI in Weather, Climate, and Coastal Oceanography (AI2ES) under NSF grant ICER-2019758. Once published, the code and data will be made available to the community via the Mountain Scholar permanent data repository with a permanent DOI and via Zenodo.


%
%

\bibliography{abstention_class_bibfile.bib}

\end{document}